\documentclass[twocolumn,showpacs,preprintnumbers,amsmath,amssymb]{revtex4}
\usepackage{graphicx}% Include figure files 
\usepackage{dcolumn}% Align table columns on decimal point 
\usepackage{bm}% bold math 
\newcommand{\eq}[1] { Eq.~(\ref{#1})}
\newcommand{\be}{\begin{equation}} 
\newcommand{\ee}{\end{equation}} 
\newcommand{\bea}{\begin{eqnarray}} 
\newcommand{\eea}{\end{eqnarray}} 
\newcommand{\av}[1] { \langle {#1} \rangle }

\begin{document}
%--------------------------------------------------------------------
\title{Anomalous Diffusion in Systems with Concentration-Dependent 
Diffusivity}
%--------------------------------------------------------------------
\author{Alex Hansen${}^{1,3,}$}
\email{alex.hansen@ntnu.no}
\author{Eirik G.\ Flekk{\o}y${}^{2,4,}$}
\email{flekkoy@fys.uio.no}
\affiliation{
${}^1$PoreLab, Department of Physics, Norwegian University of 
Science and Technology, NO--7491 Trondheim, Norway \\
${}^2$PoreLab, Department of Physics, University of Oslo, 
NO--0316 Oslo, Norway\\
${}^3$Beijing Computational Sciences Research Center, CSRC, 
10 East Xibeiwang Road, Haidian District, Beijing 100193, China\\
${}^4$PoreLab, Department of Chemistry, Norwegian University of 
Science and Technology, NO--7491 Trondheim, Norway}
%--------------------------------------------------------------------
\date{\today {}}
%--------------------------------------------------------------------
\begin{abstract}
We show analytically that there is
anomalous diffusion when the diffusion constant depends on the concentration
as a power law with a positive exponent or a negative exponent with absolute
value less than one and the initial condition is a delta function in the 
concentration.  On the other hand, when the initial concentration profile is
a step, the profile spreads as the square root of time. We verify our results
numerically using particles moving stochastically. 
\end{abstract}
%--------------------------------------------------------------------
\maketitle
%--------------------------------------------------------------------
\section{Introduction}
\label{intro}

It is often believed that the Boltzmann transformation \cite{b94} 
demonstrates that there is no anomalous diffusion when the diffusivity 
depends on the concentration. This is e.g.\ demonstrated clearly in the famous 
textbook by Crank \cite{c75}. Anomalous diffusion refers to how fast a random walker 
diffuses \cite{mk00}.  If random walker in one dimension starts at position $x=0$ when 
time $t=0$, then the RMS distance it has moved, $x_{RMS}=\sqrt{\langle x^2\rangle}$ 
when time is $t$, is
\begin{equation}
\label{eq03}
x_{RMS} \sim t^{\tau}\;.
\end{equation}
The averaging $\langle\cdots\rangle$ is done over an {\it ensemble\/} of particles.
When $\tau < 1/2$, we are dealing with {\it sub-diffusion\/} and when $1/2 < \tau \le 1$, 
we are dealing with {\it super-diffusion.\/} Normal diffusion occurs when $\tau=1/2$.

One finds a dependence of the diffusivity on concentration in many physical systems.
Newman considers examples from population dynamics and combustion \cite{n80}, Azevedo
et al.\ study water ingress in zeolites \cite{assems06,asse06}, Fischer et al.\ \cite{ffas09}
and Christov and Stone \cite{cs12} consider diffusion of grains in granular media, Hansen
et al.\ \cite{hst11} the dynamics of wetting films in wedges.  Anomalous diffusion is
reported in all of these papers.  K{\"u}ntz and Lavall{\'e}e  even gave their
paper on the diffusion of high-concentration aqueous CuSO$_4$ in deionized water
the title `Anomalous diffusion is the rule in concentration-dependent processes' \cite{kl04}.  

The diffusion equation in one dimension is
\begin{equation}
\frac{\partial}{\partial t} C(x,t)
=\frac{\partial}{\partial x} D\frac{\partial}{\partial x} C(x,t)\; ,
\label{eq01}
\end{equation}       
where $C=C(x,t)$ is
the concentration and $D$ is the diffusivity, which we in the following
assume obeys the power law
\begin{equation}
\label{eq02}
D=D(x,t)=D_0 C(x,t)^{-\gamma}\;,
\end{equation}
where $D_0$ is a constant setting the scale.  We will in the following absorb it into 
the time variable $t$.  Equation (\ref{eq01}) may then be written 
\begin{equation}
\label{eq06}
(1-\gamma)\frac{\partial}{\partial t} C(x,t)
=\frac{\partial^2}{\partial x^2} C(x,t)^{1-\gamma}\;.
\end{equation}
Hence, we see that we need $\gamma<1$ for the equation to be defined when $C(x,t)=0$.  

In the papers that assume the diffusivity to take the form (\ref{eq02}) 
\cite{n80,ffas09,hst11,cs12}, the exponent $\gamma$ is assumed to be negative.  

Pattle considered the negative-$\gamma$ case as early as 1959 \cite{p59}, indeed
finding anomalous diffusion with
\begin{equation}
\label{eq012-1}
\tau=\frac{1}{2-\gamma}\;.
\end{equation}
It is our aim here to expand the analysis of Pattle to positive $\gamma<1$ and to numerically
verify using particles that indeed the analytical solutions are the relevant ones.  One of
our major conclusions is that equation (\ref{eq012-1}) is valid for the entire range $\gamma<1$. 
As the diffusion equation is non-linear, this is not a priori given.  

We review in the next section the Boltzmann transformation and demonstrate that the initial
conditions demanded by it is a step in the concentration. In section \ref{point} we construct
the general form that the concentration profile takes.  We then go on in section \ref{gammanegative}
to consider the case when $\gamma<0$, the one studied by Pattle \cite{p59}, finding that indeed there 
is anomalous diffusion present. We present the full analytical solution here.
We then go on to section \ref{gammapositive} where we consider the $0<\gamma<1$ case.  Here, 
a full analytical solution has not been found.  However, we show that there is indeed anomalous
diffusion also in this case. We also discuss here the question of whether solutions of the 
non-linear diffusion equation are stable with respect to concentration fluctuations that are not
described by the equation.  As the equation stands, with the diffusivity given by equation (\ref{eq02}),
the solutions are not stable.  However, if we regularize the  diffusivity by adding a small
constant to it, the solutions stabilize and they describe well the process.    
Section \ref{particles} presents a numerical random walker model that
reproduces the analytical results of the previous section.  We end by summarizing our results.  

%--------------------------------------------------------------------
\section{The Boltzmann transformation and the step}
\label{boltzmann}

We assume for now that the initial  conditions is $C(x,0) = C_0
\Theta(-x)$, where $\Theta (x)$ is the Lorentz-Heaviside function.
The Boltzmann transformation consists in introducing the variable 
\begin{equation}
\label{eq02-3}
y=\frac{x}{\sqrt{t}}\; .
\end{equation}
When the $t$ and $x$ derivatives are transformed to $y$-derivatives 
the diffusion equation  (\ref{eq06}) becomes the ordinary differential equation, 
\begin{equation}
\label{eq02-5}
\frac{1-\gamma}{2}\ y \frac{dC}{dy}+\frac{d^2C^{1-\gamma}}{dy^2}=0\; 
\end{equation}
with the  $x$ and  $t$ dependence through $y$ only.
Now,  the initial condition  too
may be written in terms of $y$  alone: 
For $t=0$: $q(y<0)=C_0$ and  $q(y>0)=0$. 
For this reason the solution of the diffusion equation (\ref{eq06}) takes the form
\begin{equation}
\label{eq02-4}
C(x,t)=q(x/t^{1/2})\;
\end{equation}
for some function $q$ that satisfies \eq{eq02-5}.
This immedieately leads to the conclusion
\be  
x_{RMS}^2 = \frac{ \int dx x^2 C(x,t)  
}{\int dx  C(x,t)} \sim t, 
\ee  
i.e.  that the diffusion is normal with  $\tau=1/2$ in equation
(\ref{eq03}).
In other words, the step function initial condition cannot lead to the
anomalous diffusion behavior defined by \eq{eq03} and Pattles solution. 
In the following we shall se that this conclusion is qualitatively
changed by the introduction of a localized and thus normalizable $\delta$-function inital condition. 

%--------------------------------------------------------------------
\section{Point-like initial conditions}
\label{point}

In order to determine $\tau$ in equation (\ref{eq03}), we need need  
1.\ to specify the initial conditions so that we see how far the particles move as
time progresses. This means setting
\begin{equation}
\label{eq04}
C(x,t=0)=C_0\delta(x)\;,
\end{equation}
where $\delta(x)$ is the Dirac delta-function. We then need 2.\  to turn the concentration 
variable $C(x,t)$ into the probability density to find a particle at position $x$ 
and time $t$. This is done by normalizing $C(x,t)$, i.e.,  
\begin{equation}
\label{eq07}
\int_{-\infty}^{+\infty}dx\ C(x,t)=1\;.
\end{equation}

There is no intrinsic length or time-scale in equation (\ref{eq06}) since the
diffusivity depends on $C$ through the power law (\ref{eq02}). This means that as
long as boundary- or initial conditions do not introduce such scales
either, the solutions $C(x,t)$ must be scale-free too.
More precisely, if $x \rightarrow \lambda x$, then there must be
some rescaling of time $t \rightarrow t/f^{-1}(1/\lambda)$ so that the
probability of finding the particle remains unchanged, that is
\begin{equation}
\label{eq08-1}
dx C(x,t) = \lambda dx C\left(\lambda x, \frac{t}{f^{-1}(1/\lambda)}\right)\;.
\end{equation}
This ensures that the normalization (\ref{eq07}) remains constant with time.  We now choose
$\lambda$ so that $t/f^{-1}(1/\lambda)=1$. That is, we set 
\begin{equation}
\label{eqn08-2}
\lambda=\frac{1}{f(t)}\;.
\end{equation}
Combined with equation (\ref{eq08-1}), this gives
\begin{equation}
\label{eq08-3}
C(x,t) = \frac{1}{f(t)} C\left(\frac{x}{f(t)},1\right)=\frac{1}{f(t)} p\left(\frac{x}{f(t)}\right)\;,
\end{equation}
where we have set $p(z)\equiv C(z,1)$. 

We introduce the reduced variable
\begin{equation}
\label{eq004}
y=\frac{x}{f(t)}\;,
\end{equation}
and we have that
\begin{equation}
\label{eq005}
\frac{\partial y}{\partial x}=\frac{1}{f(t)}\;,
\end{equation}
and
\begin{equation}
\label{eq006}
\frac{\partial y}{\partial t}=-y\ \frac{\dot{f}(t)}{f(t)}\;,
\end{equation}
Equation (\ref{eq06}) may then be transformed into
\begin{equation}
\label{eq009}
\frac{1-\gamma}{2-\gamma}\ \frac{df(t)^{2-\gamma}}{dt}\ \frac{d}{dy} yp(y)+\frac{d^2}{dy^2}p(y)^{1-\gamma}=0\;.
\end{equation}

We now define
\begin{equation}
\label{eq010}
\frac{\frac{d^2}{dy^2}p(y)^{1-\gamma}}{\frac{d}{dy} yp(y)}=-c\;,
\end{equation}
giving us an equation for $f(t)$,
\begin{equation}
\label{eq011}
\frac{1-\gamma}{2-\gamma}\ \frac{df(t)^{2-\gamma}}{dt}=c\;.
\end{equation}
We integrate this equation assuming $f(0)=0$ --- since we are assuming equation
(\ref{eq04}), i.e., point-like initial conditions --- giving
\begin{equation}
\label{eq012}
f(t)=\left( \frac{2-\gamma}{1-\gamma}\ c\ t\right)^{\frac{1}{2-\gamma}}\;.
\end{equation}
This result implies that our solution takes the scaling form
\be
C(x,t) \sim \frac{g(x/t^\tau ) }{t^\tau}
\label{scal}
\ee
for some function $g$ and with $\tau $ given by \eq{eq012-1}.  
Note that this form immediately gives
\be 
x_{RMS}^2 = \frac{ \int dx x^2 C(x,t) 
}{\int dx  C(x,t)} \sim t^{2\tau} . 
\ee 

Above, we have  assumed $c >0$.  For this to be the case, using equation (\ref{eq010}),
we find that we must either have
 \begin{itemize}
  \item $d^2\ p(y)^{1-\gamma}/dy^2 < 0$ and $d(y\ p(y))/dy>0$, or\\
  \item $d^2\ p(y)^{1-\gamma}/dy^2 > 0$ and $d(y\ p(y))/dy<0$.\\
\end{itemize}
In the first case, the $p(y)^{1-\gamma}$ profile is a convex and in second case it is a concave.  
We note that a given profile $p^{1-\gamma}$ may change between being convex and concave for different values
of $y$.  That $p(y)^{1-\gamma}$ is concave or convex does not tell us whether $p(y)$ is the same.   

We note that equation (\ref{eq012}) shows that $f(t) \propto t^{1/(2-\gamma)}$. Hence, for
fixed values of $y$, i.e., for fixed values of $p(y)$, we have that $x \propto t^{1/(2-\gamma)}$. 
This is in contrast to the Boltzmann transformation, which assumes step-like initial conditions,
thus leading to $x\propto t^{1/2}$.  

%--------------------------------------------------------------------
\section{Solution for $\gamma < 0$}
\label{gammanegative}

We combine equations (\ref{eq009}) and (\ref{eq011}) to find
\begin{equation}
\label{eq013}
c\ \frac{d}{dy} yp(y)+\frac{d^2}{dy^2}p(y)^{1-\gamma}=0\;.
\end{equation}
We integrate this equation to get
\begin{equation}
\label{eq014}
c\ yp(y)+\frac{d}{dy}p(y)^{1-\gamma}=K\;,
\end{equation}
where $K$ is an integration constant. 

We now set the integration constant $K=0$ in equation (\ref{eq014}) so that we have 
\begin{equation}
\label{eq022}
c\ yp(y)+\frac{d}{dy}p(y)^{1-\gamma}=0\;.
\end{equation}
In order to non-dimensionalize this equation,
we rescale the variables $y$ and $p$,
\begin{equation}
\label{eq017}
a\tilde{y}=y\;,
\end{equation}
and
\begin{equation}
\label{eq018}
b\tilde{p}=p\;,
\end{equation} 
setting 
\begin{equation}
\label{eq023}
c a b=1\;,
\end{equation}
and
\begin{equation}
\label{eq024}
\frac{b^{1-\gamma}}{a}=1\;.
\end{equation}
Equation (\ref{eq022}) then becomes 
\begin{equation}
\label{eq025}
\tilde{y} \tilde{p}(\tilde{y})+\frac{d}{d\tilde{y}}\tilde{p}(\tilde{y})^{1-\gamma}=0\;.
\end{equation}
We see from equation (\ref{eq025}) that $d\tilde{p}(0)/d\tilde{y}\to 0$ when $\tilde{y}\to 0$.
Hence, $\tilde{p}$ approaches $\tilde{y}=0$-axis with zero slope.   

Equation (\ref{eq025}) is integrable.  We may rewrite it as
\begin{equation}
\label{eq026}
\frac{1-\gamma}{\gamma}\ d\tilde{p}(\tilde{y})^{-\gamma}=d\left(\frac{\tilde{y}^2}{2}\right)\;,
\end{equation}  
which after integration becomes
\begin{equation}
\label{eq027}
\tilde{p}(\tilde{y})=\left[\frac{1-\gamma}{2\gamma}\ (\tilde{y}_c^2-\tilde{y}^2)\right]^{-\frac{1}{\gamma}}\;,
\end{equation}
where $\tilde{y}_c^2$ is an integration constant.  
If $\gamma>0$, equation (\ref{eq027}) diverges as $|\tilde{y}|\to|\tilde{y}_c|$.  This is unphysical, and hence, 
we must have $\gamma<0$ for this solution to apply.  We combine equation (\ref{eq010}) with the 
solution (\ref{eq027}) to find
\begin{equation}
\label{eq028}
c=-2\ \frac{1-\gamma}{\gamma}\;,
\end{equation}
which is positive only when $\gamma<0$. A positive $c$ is a necessary condition for the solution to be valid.    

%---------------------------------------------- 
\begin{figure}[h!]
\begin{center}
\includegraphics[width=1.0\columnwidth]{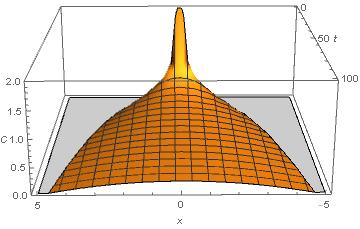}
\caption[]{The concentration field $C(x,t)$ in (\ref{eq027-1}) for $\gamma=-1$ and $y_c=1$.}
\label{fig1}
\end{center}
\end{figure}
%---------------------------------------------- 

We may now reconstruct the normalized concentration field $C(x,t)$ using equation (\ref{eq08-3}). We find
\begin{eqnarray}
\label{eq027-1}
C(x,t)&=&\Theta\left(1-\left|\frac{x}{y_ct^{\frac{1}{2-\gamma}}}\right|\right)\nonumber\\ 
&&\frac{1}{y_c t^{\frac{1}{2-\gamma}}}\ 
\frac{\Gamma\left[\frac{3}{2}-\frac{1}{\gamma}\right]}{\sqrt{\pi}
\Gamma\left[1-\frac{1}{\gamma}\right]} \left[1-\left(\frac{x}{y_ct^{\frac{1}{2-\gamma}}}\right)^2\right]^{-\frac{1}{\gamma}}\;,
\nonumber\\
\end{eqnarray}
where $y_c=a\tilde{y}_c$.  This is the solution that was found by Pattle \cite{p59}.  We show it in figure \ref{fig1}
for $\gamma=-1$ and $y_c=1$. 

We now calculate $x_{RMS}$ for $\gamma<0$. Hence,    
\begin{equation}
\label{eq027-2}
x_{RMS}^2=\int_{-y_ct^{1/(2-\gamma)}}^{+y_ct^{1/(2-\gamma)}} dx\ x^2 C(x,t)=\frac{-\gamma}{2-3\gamma} y_ct^{\frac{2}{2-\gamma}}\;,
\end{equation}
where $C(x,t)$ is given by the solution (\ref{eq027-1}). By comparing with equation (\ref{eq03}) we find
equation (\ref{eq012-1}) --- as did Pattle. 

%--------------------------------------------------------------------
\section{Solution for $0<\gamma < 1$}
\label{gammapositive}

We return to equation (\ref{eq014}), now assuming that $K\neq 0$. 
We divide the equation by $|K|$ to get
\begin{equation}
\label{eq015}
\frac{c}{|K|}\ yp(y)+\frac{1}{|K|}\ \frac{d}{dy}p(y)^{1-\gamma}=\epsilon\;,
\end{equation}
where
\begin{equation}
\label{eq016}
\epsilon=\frac{K}{|K|}=\pm 1\;.
\end{equation}
This equation cannot be integrated directly as could the case for $K=0$. However,
we will be able to pry the essential information from it anyway. 

We non-dimensionalize equation (\ref{eq015}) by invoking
equations (\ref{eq017}) and (\ref{eq018}) and setting 
\begin{equation}
\label{eq019}
\frac{c}{|K|}\ a b=1\;,
\end{equation}
and
\begin{equation}
\label{eq020}
\frac{1}{|K|}\ \frac{b^{1-\gamma}}{a}=1\;.
\end{equation}
Equation (\ref{eq015}) thus becomes
\begin{equation}
\label{eq021}
\tilde{y} \tilde{p}(\tilde{y})+\frac{d}{d\tilde{y}}\tilde{p}(\tilde{y})^{1-\gamma}=\epsilon\;.
\end{equation}
If $\epsilon=+1$, we must have from this equation that $d\tilde{p}(0)^{1-\gamma}/d\tilde{y}=1$ when $\tilde{y}=0$. 
A positive derivative at the origin means that $\tilde{p}(\tilde{y})$ {\it increases\/} 
as we move away from the origin.  We may therefore discard this possibility as being unphysical. 
On the other hand, if $\epsilon=-1$, we must have that $d\tilde{p}(0)^{1-\gamma}/d\tilde{y}=-1$ 
when $\tilde{y}=0$, which makes physically sense.  Hence, only the $\epsilon=-1$ case needs to be
pursued further.

Suppose now that $\tilde{p}(\tilde{y})>0$ for all finite $\tilde{y}$.  Since $C(x,t)$ is normalizable, we know that
$\tilde{p}(\tilde{y})\to 0$ as $|\tilde{y}|\to\infty$ faster than $1/|\tilde{y}|$.  Hence, $\tilde{y}\tilde{p}(\tilde{y})\to 0$
as $|\tilde{y}|\to\infty$.  From equation (\ref{eq021}) we then have that   
\begin{equation}
\label{eq029-1}
\frac{d}{d\tilde{y}}\ \tilde{p}(\tilde{y})^{1-\gamma}\to -1\quad {\rm as}\ |\tilde{y}|\to\infty\;.
\end{equation}
This is not possible, and we conclude that there is a {\it finite\/} $\tilde{y}_c$ such that $\tilde{p}(\tilde{y})=0$ for  
$|\tilde{y}|\ge\tilde{y}_c$. 

We will in the following investigate how the solution of equation (\ref{eq021}) behaves close to $\tilde{y} \approx 0$ and
$\tilde {y}\approx \tilde{y}_c>0$.  
At $\tilde{y}=\tilde{y}_c$,
equation (\ref{eq021}) becomes
\begin{equation}
\label{eq029}
\frac{d}{d\tilde{y}}\ \tilde{p}(\tilde{y})^{1-\gamma}=-1\;.
\end{equation}
We integrate this equation to find
\begin{equation}
\label{eq030}
\tilde{p}(\tilde{y})=(\tilde{y}_c-\tilde{y})^{\frac{1}{1-\gamma}}\;.
\end{equation}
This is the lowest order expansion of $\tilde{p}$ around $\tilde{y}_c$.  In order to find the next order, we assume
$\tilde{p}$ to take the form
\begin{equation}
\label{eq031}
\tilde{p}(\tilde{y})^{1-\gamma}=(y_c-y)+\delta\tilde{p}(\tilde{y})\;.
\end{equation}
We insert this expression into equation (\ref{eq021}) and find
that $\delta\tilde{p}(\tilde{y})$ obeys the equation
\begin{equation}
\label{eq032}
\tilde{y}_c(\tilde{y}_c-\tilde{y})^{\frac{1}{1-\gamma}}
+\frac{d}{d\tilde{y}}\ \delta\tilde{p}(\tilde{y})=0\;.
\end{equation}
We solve this equation and find
\begin{equation}
\label{eq033}
\delta\tilde{p}(\tilde{y})=\frac{1-\gamma}{2-\gamma}\tilde{y}_c(\tilde{y}_c-\tilde{y})^{\frac{2-\gamma}{1-\gamma}}\;.
\end{equation}
Combining this result with equation (\ref{eq031}) gives
\begin{equation}
\label{eq034}
\tilde{p}(\tilde{y})=\left(\tilde{y}_c-\tilde{y}\right)^{\frac{1}{1-\gamma}}
\left[1+\frac{\tilde{y}_c}{2-\gamma}(\tilde{y}_c-\tilde{y})^{\frac{1}{1-\gamma}}+\cdots\right]\;.
\end{equation}

We see from this expression that $d\tilde{p}(\tilde{y})/d\tilde{y}\to 0$ as $\tilde{y}\to\tilde{y}_c$, i.e., 
the profile approaches the maximum $\tilde{y}$ value with a slope that goes to zero. However, we have that
\begin{equation}
\label{eq034-1}
\frac{d^2\tilde{p}(\tilde{y})}{d\tilde{y}^2}=\frac{\gamma}{(1-\gamma)^2} (\tilde{y}_c-\tilde{y})^{\frac{2\gamma-1}{1-\gamma}}\;.
\end{equation}
This expression is always positive and the $\tilde{p}(\tilde{y})$ profile is therefore always concave. However,
we note that for $0<\gamma<1/2$, the second derivative diverges. Hence, the first derivative reaches zero in a `brutal' way
for these values of $\gamma$.

We now use equation (\ref{eq010}) combined with the equations (\ref{eq033}) and (\ref{eq034})
to find
\begin{equation}
\label{eq035}
c=1\;,
\end{equation}
to lowest order in $\tilde{y}_c-\tilde{y}$. Hence, $c>0$ and the solution for $\epsilon=-1$ is viable.  

From equations (\ref{eq021}) and (\ref{eq029}), we have that
\begin{equation}
\label{eq036}
\frac{d}{d\tilde{y}}\ \tilde{p}(\tilde{y})^{1-\gamma}\to -1 \quad {\rm as\ } \tilde{y}\to 0\ {\rm or\ } 
\tilde{y}\to \tilde{y}_c\;.
\end{equation}
Hence, $\tilde{p}(\tilde{y})$ leaves the $\tilde{p}$ axis at $\tilde{y}=0$ with the same slope as it reaches the $\tilde{y}$ 
at $\tilde{y}_c$.  

At $\tilde{y}=0$, we have that 
\begin{equation}
\label{eq034-2}
\frac{d\tilde{p}(y)^{1-\gamma}}{d\tilde{y}}=-1\;
\end{equation}
to lowest order in $\tilde{y}$.  We integrate this expression and find
\begin{equation}
\label{eq034-3}
\tilde{p}(\tilde{y})^{1-\gamma}=\tilde{p}(0)^{1-\gamma}-\tilde{y}\;,
\end{equation}
which to lowest order in $\tilde{y}$ gives
\begin{equation}
\label{eq034-4}
\tilde{p}(\tilde{y})=\tilde{p}(0)-\frac{\tilde{p}(0)^\gamma}{1-\gamma}\ y\;.
\end{equation}
We see that $\tilde{p}(\tilde{y})$ approaches the $\tilde{y}=0$-axis at an angle. Hence, the maximum of the of the concentration
profile forms a wedge. 
 
Furthermore, for small $\tilde{y}>0$, we have that $d\tilde{y}\tilde{p}(\tilde{y})/d\tilde{y}=\tilde{p}(\tilde{y})+
\tilde{y}d\tilde{p}(\tilde{y})/d\tilde{y}>0$ since we can make the second term in the middle as small as we wish by
making $\tilde{y}$ small enough. Hence, we must have
$d^2\ \tilde{p}(\tilde{y})^{1-\gamma}/d\tilde{y}^2 < 0$ to ensure $c>0$.  This must be the case 
in order for the two terms on the left hand side in equation (\ref{eq013}) to sum to zero. So,
$\tilde{p}(\tilde{y})^{1-\gamma}$ must be convex near the origin. Since $1-\gamma<0$, $\tilde{p}(\tilde{y})$
must also be concave near $\tilde{y}=0$. 

Near $\tilde{y}_c$, we have that
\begin{equation}
\label{eq037}
\frac{d}{d\tilde{y}}\ \tilde{y}\tilde{p}(\tilde{y})=-\frac{y_c}{1-\gamma}\ 
(\tilde{y}_c-\tilde{y})^{\frac{\gamma}{1-\gamma}}<0\;.
\end{equation}
We furthermore find
\begin{equation}
\label{eq038}
\frac{d^2}{d\tilde{y}^2}\ \tilde{p}(\tilde{y})^{1-\gamma}=\frac{\tilde{y}_c}{1-\gamma}\
(\tilde{y}_c-\tilde{y})^{\frac{\gamma}{1-\gamma}}>0\;.
\end{equation}
That is, $\tilde{p}(\tilde{y})^{1-\gamma}$ is concave near $\tilde{y}_c$.  Since $1-\gamma<0$, $\tilde{p}(\tilde{y})$
must also be concave near $\tilde{y}_c$. 
Furthermore, we see that equation (\ref{eq038}) diverges if $\gamma<0$ and it is well behaved if 
$\gamma>0$.  Our conclusion is that the $\epsilon=-1$ solution corresponds to $\gamma>0$.

If we had the exact profile $\tilde{p}(\tilde{y})$, we could have proceeded to construct the normalized
concentration field $C(x,t)$ as we did in equation (\ref{eq027-1}) for $\gamma<0$.  We do not have this profile,
but we may still conclude that equation (\ref{eq012-1}) works also for $0<\gamma<1$, since 
\begin{eqnarray}
\label{eq027-5}
x_{RMS}^2&=&\int_{-y_ct^{1/(2-\gamma)}}^{+y_ct^{1/(2-\gamma)}} dx\ x^2 C(x,t)\nonumber\\
&=&\int_{-y_ct^{1/(2-\gamma)}}^{+y_ct^{1/(2-\gamma)}} 
dx\ \frac{x^2}{t^{\frac{1}{1-\gamma}}}p\left(\frac{x}{t^{\frac{1}{1-\gamma}}}\right)\nonumber\\
&\propto& t^{\frac{2}{1-\gamma}}\;.
\end{eqnarray}

%--------------------------------------------------------------------
\subsection{Does the $0<\gamma<1$ solution really exist?}
\label{skrekkoggru}

When $0<\gamma<1$, the diffusivity given by equation (\ref{eq02}) diverges. Still,
the non-linear diffusion equation (\ref{eq06}) is well behaved and has solutions,
even if we are unable to write them down explicitly.  We will in the following
section model the diffusion processes described by (\ref{eq06}) by a stochastic process
involving diffusing particles.  However, let us forego this discussion and already now
picture the diffusion process described by (\ref{eq06}) with $0<\gamma<1$.  Focus on
the region close to but to the left of the sharp front at $+\tilde{y}_c$.  This region will be 
swarming with particles.  There will always be a particles which is furthest to the 
right. This particle will be alone.  Hence, according to the diverging diffusivity, this
particle will be kicked off to $x\to\pm\infty$ and be gone.  Then, there will be another particle
furthest to the right which therefore will be alone, and the same happens to this one.  And
so on.  Soon there will be no particles left. 

This leakage is caused by fluctuations that are not described by the diffusion equation. In
this case, they must dominate the process and they have a devastating effect on the solution 
of the non-linear diffusion equation we have just described.  The solution to this dilemma
is to add a small positive number $\delta$ to the concentration in this equation so that 
it becomes
\begin{equation}
\label{eq02-101}   
D=D_0\left[C(x,t)+\delta\right]^{-\gamma}\;.
\end{equation}
This changes the character of the diffusion equation when $C\approx \delta$, but it 
stops the ``leakage" due to fluctuations.  From a physical point of view, it makes sense that
the the diffusion process goes normal for small enough concentrations. When  $\delta$ is added
in equation (\ref{eq02-101}), the solution we have described here is still valid for $C>\delta$.  
We demonstrate this numerically in Section \ref{positive}. Hence, the approach taken in this
section is physically realistic.  

%--------------------------------------------------------------------
\section{Derivation of the diffusion equation from particle dynamics}
\label{particles}

We now turn to stochastic modeling of the process we so far have described using
the non-linear diffusion equation (\ref{eq06}).   

%----------------------------------------------
\begin{figure}[h!]
\begin{center}
\includegraphics[width=1.00\columnwidth]{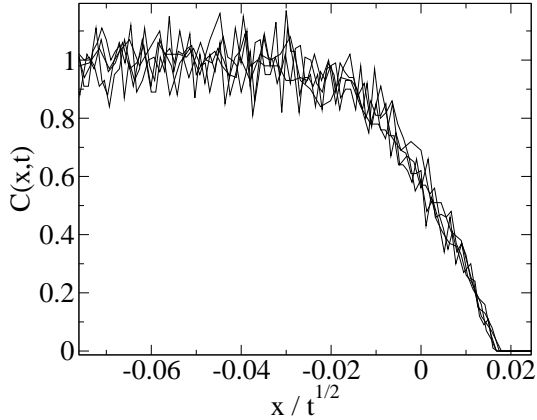}
\caption[]{The concentration $C(x,t)$ as a function of  
$x/t^{1/2}$ when $\gamma=-1$ and the initial concentration was
$C(x,0)=\Theta(-x)$, the Heaviside step function.}
\label{fig2}
\end{center}
\end{figure} 
%---------------------------------------------- 

Following the discussion in the van Kampen book {\it Stochastic Processes in
Physics and Chemistry\/} \cite{k07}, we derive the non-linear diffusion 
equation, which is an example of a Fokker-Planck equation from the
following particle model:
A population of $N_p$ particles are propagated by a sequence of
random steps of zero mean using a {\it concentration dependent
step length.\/}  For every time the particle positions, which take on
continuous values,  are updated, the
concentration field $C(x,t)$ is updated onto a discrete one-dimensional lattice of
unit lattice constant. 
Their positions $x_i$  are updated
according to the following algorithm
\begin{equation}
x \rightarrow x +\Delta x\;, 
\label{udate}
\end{equation}  
where
\begin{equation}
\Delta x = \eta g(C(x)) \sqrt{\Delta t}
\label{kjbyf}
\end{equation}
is a Wiener process and $\eta$ is a random variable with
$\av{\eta}=0$ and  $\av{\eta^2}=1$, where  $g(C)$ may be the
normalized concentration.
Now, following the discussion in the van Kampen  book, Chapter
VIII.2  we derive the corresponding Fokker-Planck equation as follows.
The Chapman-Kolmogorov or master equation describing the above stochastic process is
\begin{equation} 
\frac{\partial C(x,t) }{\partial t }
= \int_{-\infty}^\infty dr C(x-r, t) W(x-r,r) - C(x, t) W(x,-r)\;,
\label{eq055-1}
\end{equation}
where $W(x,r)$ is the number  of particles  per time and length that jump a distance $r$
starting from $x$.  We note that the formalism remains valid also when $W(x,t)$ depends on $x$
via $c(x,t)$ itself. Taylor expanding the integrand around $x$ yields the Fokker-Planck equation
\begin{equation}
\frac{\partial C(x,t) }{\partial t }
=
\frac{1}{2}\frac{\partial^2 }{\partial x^2}\left( a_2(x) C(x,t)
\right)\;,
\label{hkuft}
\end{equation}
where $a_2(x)$ is the mean squared jump length per time,
\begin{equation}
a_2(x) = \int dr r^2 W(x,r) =\frac{\av{\Delta x^2}}{\Delta t} = g(C)^2\;,
\end{equation}
according to equation (\ref{kjbyf}). Setting $g(C) = b C^{-\gamma /2}$ gives
\begin{equation}
\frac{\partial C}{\partial t}
= \frac{b^2}{2} 
\frac{\partial^2 }{\partial x^2} C^{1-\gamma}\;, 
\label{eq055-2}
\end{equation}
and requiring equivalence with 
equation (\ref{eq06}) thus implies that $b^2 = 2/(1-\gamma )$.

%-------------------------------------------------------
\subsection{It{\^o}-Stratonovitch dilemma}
\label{ito}

However, the presence of a $C$-dependence in the diffusivity $D$ 
introduces an ambiguity in the implementation of equation (\ref{udate}), 
since now $\Delta x$ also depends on $C$, which in turn depends on all the 
$\Delta x$'s. So, the question is whether one should use $C(x)$ or $C(x + \Delta x)$
or perhaps something in between? Since $\Delta x \sim \sqrt{\Delta t}$
these choices are not equivalent.

%---------------------------------------------- 
\begin{figure}[h!]
\begin{center}
\includegraphics[width=1.00\columnwidth]{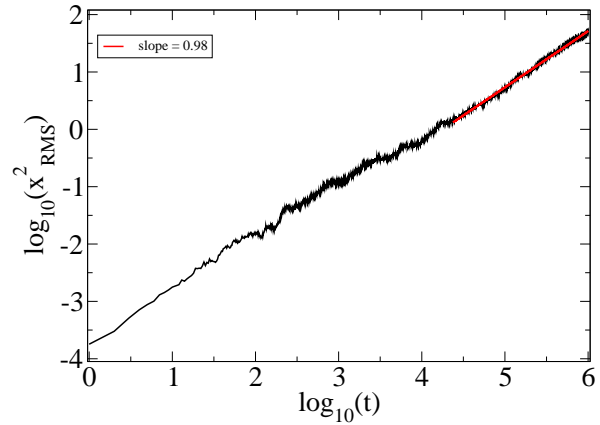}
\caption[]{The mean square displacement as a function of time when
$\gamma=-1$ and initial condition $C(x,0)=\Theta(-x)$, the Heaviside 
step function. According to the Boltzmann transformation, we expect
$x_{RMS} \sim t^{1/2}$, which is precisely what we find.}
\label{fig3}
\end{center}
\end{figure}
%---------------------------------------------- 

Stratonovitch read equation (\ref{udate}) as  \cite{k07}
\begin{equation}  
x \rightarrow x + g\left( \frac{c(x(t + \Delta t) ) + C(x(t)  )
  }{2}\right) \eta \sqrt{\Delta t}\;,
\label{udates}
\end{equation}  
while It{\^o} read it as 
\begin{equation} 
x \rightarrow x + g\left( C\left(x(t) \right)\right) \eta \sqrt{\Delta t}\;.
\label{udatei}
\end{equation}
It turns out that it is the choice  opted for by It{\^o} that gives
equation (\ref{hkuft}), while the Stratonovitch choice gives 
\begin{equation}  
\frac{\partial C}{\partial t}
= \frac{1}{2} \frac{\partial }{\partial x}\left(  g \frac{\partial  
  }{\partial x}\left( g C \right) \right)\;, 
\end{equation}  
see van Kampen's book \cite{k07}, Chapter VI.4 for a derivation of this result.
By setting $g(C) = b C^{-\gamma /2}$ again we can write the above equation as
\begin{equation} 
\frac{\partial C}{\partial t}
= \frac{b^2}{2} \frac{1-\gamma /2}{1 -\gamma }
\frac{\partial^2 }{\partial x^2} C^{1-\gamma}\;, 
\end{equation} 
and equivalence with  
equation (\ref{eq06}) now implies that $b^2 = 2/(1-\gamma /2)$.
This means that the only difference between the
It{\^o} and Stratonovitch implementations of equation (\ref{udate})
is the magnitude of the random step. In the It{\^o} case the step length
has to be a bit smaller than in the Stratonovitch case in order to
correspond to the same macroscopic descriptions for $\gamma\neq 0$. 
When $\gamma =0$ the $C-$-dependence of $D$ goes away and the two
interpretations give the same $b$, as one would expect. 

%----------------------------------------------
\begin{figure}[h!]
\begin{center}
\includegraphics[width=1.00\columnwidth]{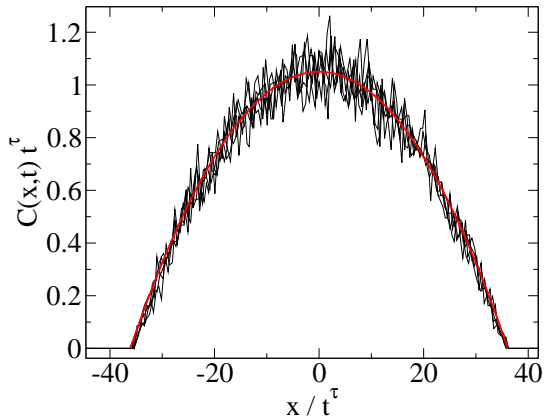}
\caption[]{The scaled concentration $t^\tau C(x,t)$ as a function of  
$x/t^{\tau} $where $\tau=1/3$ 
according to equation (\ref{eq012-1}) when $\gamma=-1$ and $dt =5\; 10^{-5}$.
The black lines show simulations at different times  
while the red shows a parabolic fit.}
\label{fig4}
\end{center}
\end{figure} 
%---------------------------------------------- 

In the simulations it is convenient to use the It{\^o} implementation
and thus $\Delta x = \eta \sqrt{2\Delta t/(1-\gamma )}c^{-\gamma /2}$.
First, the particles are 
initialized at the same location, so that the initial concentration is 
a $\delta$-function. The time step is $dt =5\; 10^{-5}$ and $N_p=1000$
particles are used.  

%---------------------------------------------- 
\begin{figure}[h!]
\begin{center}
\includegraphics[width=1.00\columnwidth]{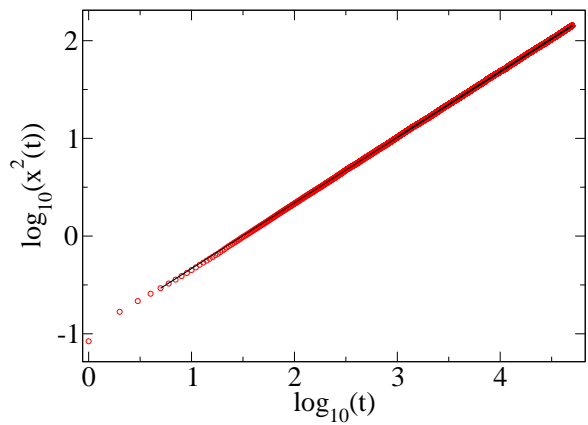}
\caption[]{The mean square displacement as a function of time when
$\gamma=-1$ so that $\tau=1/3$ according to equation (\ref{eq012-1}).
Hence, we expect $x_{RMS} \sim t^{2/3}$.}
\label{fig5}
\end{center}
\end{figure}
%---------------------------------------------- 

%-------------------------------------------------------
\subsection{The concentration is initially a step function}
\label{boltzmann-2}

We start by considering the step initial conditions first studied by
Boltzmann \cite{b94}.  We start the simulations by setting $C(x,t=0)=\Theta(-x)$,
where $\Theta(x)$ is the Heaviside step function, which is 0 for negative arguments
and one for positive arguments.  We show in figure \ref{fig2} the concentration 
profile $C(x,t)$ for different times plotted against the reduced
variable $y=x^/\sqrt{t}$, see equation (\ref{eq02-3}) using $\gamma=-1$.  
There is data collapse in accordance with equation (\ref{eq02-4}).

In figure \ref{fig3} we show the RMS displacement $\sum_{i=1}^N
x_i^2/N$ where the sum runs over all particles with positions $x_i>0$.
 This quantity is easily calculated as the motion of each
particle is traced.

%-------------------------------------------------------
\subsection{Delta-function initialization: the $\gamma<0$ case}
\label{negative}

In this case we chose $\gamma=-1$. We had all the particles collected at the 
origin for $t=0$, thus fulfilling the initial condition (\ref{eq04}).  We then let the
particles loose with the result shown in figure \ref{fig4}: $t^\tau C(x,t)$ plotted 
against $x/t^\tau$ where $\tau$ is given by equation (\ref{eq012-1}), and hence equal to
$1/3$.  We have the exact solution for $C(x,t)$ for negative $\gamma$ given in equation 
(\ref{eq027-1}). When $\gamma=-1$, we expect a parabolic shape.  We show this parabola in
red in figure \ref{fig4}.  

Figure \ref{fig5} shows $x_{RMS}^2$ vs.\ $t$ on log-log scale.  The straight line is a fit
and we measure $\tau=0.67$  in comparison to the theoretical value
$2\tau=2/3$, see equation (\ref{eq027-2}).  

%-------------------------------------------------------
\subsection{Delta-function initialization: the $0<\gamma<1$ case}
\label{positive}

Here we set $\gamma=1/2$. We use the regularized diffusivity given in equation (\ref{eq02-101})
with $\delta=0.01$.  

%----------------------------------------------
\begin{figure}[h!]
\begin{center}
\includegraphics[width=1.00\columnwidth]{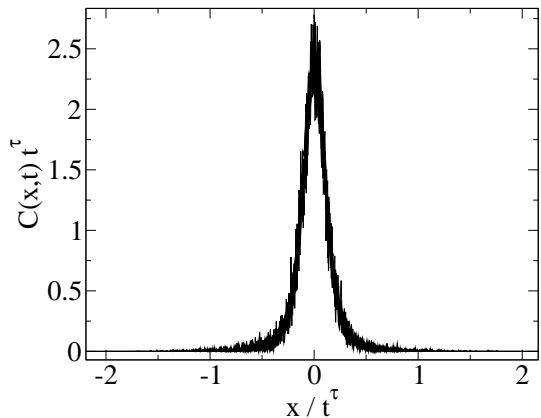}
\caption[]{The scaled concentration $t^\tau C(x,t)$ as a function of  
$x/t^{\tau}$ when $\gamma=1/2$ so that $\tau=2/3$ according to equation (\ref{eq012-1}).}
\label{fig6}
\end{center}
\end{figure}
%----------------------------------------------

We had as in the negative-$\gamma$ case all the particles 
collected at the origin for $t=0$, in accordance with the initial condition (\ref{eq04}).  
The ensuing result is shown in figure \ref{fig6}: $t^\tau C(x,t)$ plotted 
against $x/t^\tau$ where $\tau$ is given by equation (\ref{eq012-1}). In this case it is
$2/3$.  We do not have the analytical form of the profile in contrast to the negative $\gamma$
case.

\begin{figure}[h!]
\begin{center}
\includegraphics[width=1.00\columnwidth]{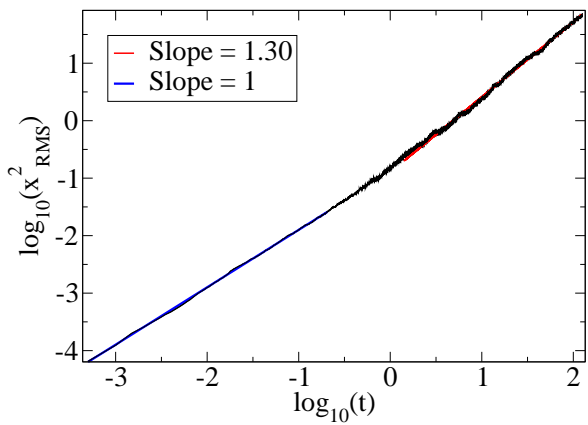}
\caption[]{The mean square displacement as a function of time when
$\gamma=1/2$ so that $\tau=2/3$ according to equation (\ref{eq012-1}).
Hence, we expect $x_{RMS}^2 \sim t^{4/3}$.  The black line is a 
fit with exponent $1.30$.}
\label{fig7}
\end{center}
\end{figure}

Figure \ref{fig7} shows $x_{RMS}^2$ vs.\ $t$ on log-log scale.  The straight line is a fit
and we measure $\tau=0.67$ in good comparison to the theoretical value
$2\tau=4/3$.   

%---------------------------------------------- 
%---------------------------------------------- 

%--------------------------------------------------------------------
\subsection{Step-function initialization: the $0<\gamma<1$ case}
\label{negativestep}

\begin{figure}[h!]
\begin{center}
\includegraphics[width=1.00\columnwidth]{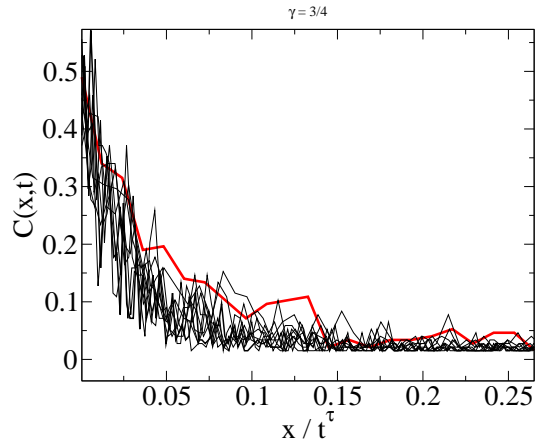}
\includegraphics[width=1.00\columnwidth]{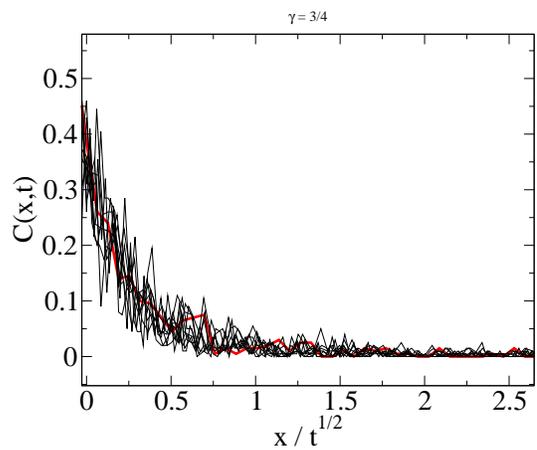}
\caption[]{
The evolution of the step profile when $\gamma=3/4$. We have plotted it as a function of
position scaled by the two different factors $t^{\tau}$ where $\tau=4/5$ according to equation 
(\ref{eq012-1}), and $t^{1/2}$. The first profile is red, the latter black.}
\label{figstep}
\end{center}
\end{figure}

As a test of the step function behavior for $\gamma >$0 we set $\gamma
=3/4$ and initialize $N_p=$ 8000 particles at a constant density in a
region $x\le 0$. The results are shown in figures \ref{figstep} where we
plot $C(x,t )$ against both $x/t^{1/2}$ and $x/t^{\tau}$.

%---------------------------------------------- 
%---------------------------------------------- 

\begin{figure}[h!]
\begin{center}
\includegraphics[width=1.00\columnwidth]{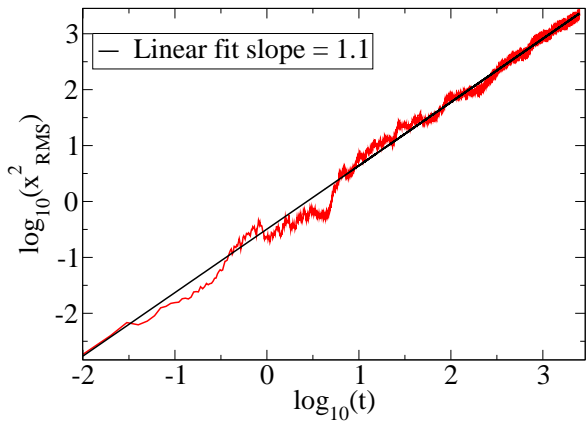}
\caption[]{The mean square displacement $\int_0^\infty dx x^2 C(x,t) /
  \int_0^\infty dx  C(x,t) $ when $\gamma =  3/4$.}
\label{figstepx2}
\end{center}
\end{figure}

It is seen that the data-collapse is somewhat better for  the $x/t^{1/2}$ choice.
Likewise, figure \ref{figstepx2} shows a clear normal-diffusion scaling
of $\av{x^2}$. 

Hence, also in the $0<\gamma<1$ case, we have that the step profile leads to normal diffusion. 

%---------------------------------------------- 
%---------------------------------------------- 

\begin{figure}[h!]
\begin{center}
\includegraphics[width=1.00\columnwidth]{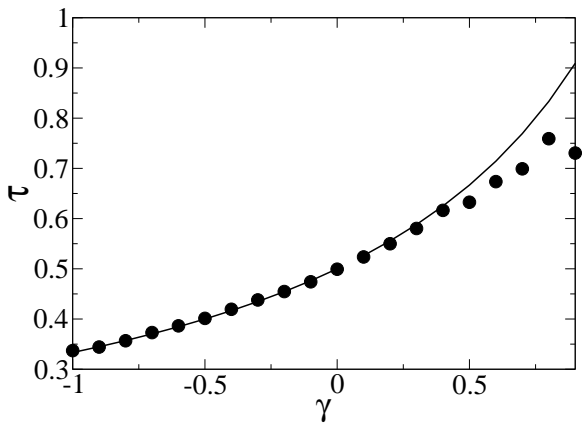}
\caption[]{The measured diffusion exponent ($\bullet$) compared to the
prediction of equation (\ref{eq012-1}) (full line).}
\label{fig8}
\end{center}
\end{figure}

\bigskip
We summarize our numerical findings in figure \ref{fig8} where we
compare the measured values of $\tau$ compared to the prediction in
equation (\ref{eq012-1}) over the range of $\gamma$-values $[-1,1]$,
$\gamma=0$ excluded.  As is apparent, the coincidence between the
prediction (\ref{eq012-1}) and the measured values are decreasingly 
matching as $\gamma$ approaches 1.  There are two reasons for this,
the first one being that the singularity in the diffusivity, equation (\ref{eq02}),
becomes more severe with increasing $\gamma$. The second reason is that
the coupling between the master equation (\ref{eq055-1}) and the Fokker-Planck
equation (\ref{eq055-2}) becomes increasingly tenuous as the expansion is
done around a singular point.   

%---------------------------------------------- 
%---------------------------------------------- 

%--------------------------------------------------------------------
\section{Summary and conclusions}
\label{conclusions}

A power law dependence of the diffusivity with respect to the concentration, 
equation (\ref{eq02}) leads to anomalous diffusion.  That is, the root-mean-square
distance moved by a particle does not scale as the square root of time,
but another power, see equation (\ref{eq03}).  The way to measure this quantity
using the concentration field is to initialize the system with a delta function in
the concentration.  The result of Boltzmann \cite{b94} going back 125 years, is still 
surprising in light of this.  When the concentration is initiated as a step function,
the anomalous behavior seems to disappear: if we follow a given level of concentration 
$C(x(t),t)$ in time, we find $x(t)\sim \sqrt{t}$ in this case, as in normal
diffusion.  

With a power law diffusivity and a delta-function initial condition,
there is no length scale in the problem.
In this case normalizability leads to the scaling form of
equation (\ref{scal})
leading to  the exponent relation equation (\ref{eq012-1})
which is the defining characteristic of anomalous diffusion.
With a step function initial condition however, the solution extends
to $x=- \infty$  and cannot any more by normalized.
Hence, in this case, equation (\ref{scal}) is replaced by equation (\ref{eq02-4})
which gives  normal $\tau = 1/2$ diffusion.
If the step function were modified to a normalizable profile, it would necessarily imply the introduction 
of a length scale. 

We have in this paper reviewed the Boltzmann result and demonstrated, as did Pattle
60 years ago \cite{p59}, that when $D(C)\sim C^{-\gamma}$ where $\gamma<0$, the
non-linear diffusion equation is analytically solvable and indeed leads to anomalous
diffusion.  We go on, however, to consider the case when $0<\gamma<1$ which is not 
analytically solvable.  Also this case shows anomalous diffusion, and we work this out
analytically even though we are not able to solve for the entire concentration profile.

We then go on to construct a stochastic particle dynamics that we implement computationally.
Using this approach, we are able to verify the central results we have derived earlier.
They all match. 

A couple of remarks at the very end:       

K{\"u}ntz and Lavall{\'e}e \cite{kl04} conclude their abstract of their paper 
with the words `Spreading fronts are subdiffusive for $D(C)$ decreasing with $C$,
superdiffusive for increasing $D(C)$ and scale only as $t^{1/2}$ only for
constant $D$.' Our findings here are the opposite.  We find superdiffusive behavior
when $<0\gamma<1$, i.e., $D(C)$ decreasing with increasing $C$ and we find 
subdiffusive behavior when $\gamma<0$, i.e., $D(C)$ increasing with increasing $C$.
However, if we compare figures \ref{fig4} and \ref{fig6} where we plot $t^\tau C(x,t)$ 
against $x/t^\tau$ for $\gamma=-1<0$ and for $\gamma=1/2>0$, we see that the former curve
($\gamma<0$) which is a parabola, is `fatter' than the latter curve ($\gamma>0$), which
gives the appearance of a `skinny' bell curve.  Hence, relatively rather than in absolute
terms, the $\gamma<0$ case propagates the walkers further away from the origin than
the $\gamma>0$.  In this sense, we agree with K{\"u}ntz and Lavall{\'e}e.

We mentioned in the introduction, anomalous diffusion originating from a concentration
dependent diffusivity may have been seen in diffusion in granular media \cite{ffas09,cs12}.
These observations are based on rotating a bi-disperse composition of smaller and large glass
beads in a horizontal cylindrical mixer.  The mixer is filled with the larger beads except
for a small disk of smaller beads.  As the cylinder turns, the smaller beads diffuse into the
larger beads and the concentration of smaller beads as a function of time and position along
the cylinder is recorded.  This setup mimics closely the initial conditions that we have studied
here, except for Section \ref{boltzmann}, where we assumed a step initially. The connection with the 
present work is the proposal that the diffusivity of the smaller beads is larger when they are
surrounded by other smaller beads than when they are surrounded by the larger beads; the higher the
concentration of smaller beads, the larger their diffusivity is.  We propose here to prepare the
packing in a different way initially.  Fill (say) the left half of the cylinder with the smaller
beads and the right half with the larger beads.  The system is therefore initiated with a step 
function in the concentration.  According to Boltzmann, as demonstrated in Section \ref{boltzmann}, 
one would then expect {\it normal diffusion\/} where the front evolves as $x^2\sim t$, i.e., the 
parabolic law.  

%--------------------------------------------------------------------
\begin{acknowledgements}
This work was partly supported by the
Research Council of Norway through its Centers of Excellence funding
scheme, project number 262644.  We thank M.\ R.\ Geiker, P.\ McDonald and
members of the ERICA network for interesting discussions. AH thanks
Hai-Qing Lin and the CSRC for friendly hospitality.   
\end{acknowledgements}

%--------------------------------------------------------------------

%--------------------------------------------------------------------
\end{document}